\documentclass[twocolumn,aps,prl,showpacs]{revtex4}
\usepackage{amsmath}
\usepackage{graphicx}
\usepackage{dcolumn}
\usepackage{bm}
\usepackage{float}
\usepackage{xcolor}

\usepackage{amssymb}
\usepackage{txfonts}
\usepackage{wasysym}

\definecolor{ao(english)}{rgb}{0.0, 0.5, 0.0}

\begin{document}

\draft
\emph{}
\title{Aging of the Linear Viscoelasticity of Glass- and Gel-forming Liquids}
\author{O. Joaqu\'in-Jaime$^{1}$, E. Lázaro-Lázaro$^{1}$, R. Peredo-Ortiz$^{1}$, \\ 
S. Srivastava$^{2}$, M. Medina-Noyola$^{1}$ and L.F. Elizondo-Aguilera$^{3}$}
\affiliation{$^{1}$ Instituto de F\'{\i}sica,
Universidad Aut\'{o}noma de San Luis Potos\'{\i}, \'{A}lvaro
Obreg\'{o}n 64, 78000 San Luis Potos\'{\i}, SLP, M\'{e}xico}
%\affiliation{$^{2}$ Facultad de Ciencias F\'isico-Matem\'aticas,
%Benem\'erita Universidad Aut\'{o}noma de Puebla, Apartado Postal 
%1152, CP 72570, Puebla, PUE, M\'{e}xico}
\affiliation{$^{2}$ Department of Chemical  and Biomolecular Engineering, University of California, Los Angeles, Los Angeles, California  90095, United  States}
\affiliation{$^3$ Instituto de F\'isica, Benem\'erita Universidad Aut\'onoma de Puebla, 
Apartado Postal J-48, 72570 Puebla, Mexico.}
\
\date{\today}

\begin{abstract}

We report a novel approach based on the non-equilibrium self-consistent generalized Langevin equation (NESCGLE) theory that allows for the first principles prediction of the zero-shear viscosity in glass- and- gel-forming materials. This new modulus of the NESCGLE theory facilitates the theoretical description and interpretation of experimental data concerning out-of-equilibrium rheological properties of viscous liquids during their amorphous solidification. The predictive capability of our approach is illustrated here by means of a quantitative comparison between theoretical and experimental results for the zero shear viscosity in suspensions of oligomer-tethered nanoparticles in a polymeric host, finding an almost perfect correspondence between experiments and theory. This comparison also highlights the crucial relevance of including a kinetic perspective, such as that provided by the NESCGLE theory, in the description of dynamic and viscoelastic properties of amorphous states of matter.

\end{abstract}

%\pacs{05.70.Ln, 64.70.P−, 66.20.Cy}

\maketitle

%%%%%%%%
\section{Introduction}\label{sec1}
%%%%%%%%

Aging is an essential fingerprint of the process of amorphous solidification of glass- and gel-forming liquids \cite{cipellettiramos05,cipelletti2,lunkenheimer,di,bandyopadhyay,bissing,hecksher}. It is consistently observed in a wide variety of microscopically very different materials. These include metallic alloys \cite{das,ruta},  polymers \cite{hutchinson,bellon}, polymeric solutions \cite{baudez01}, colloidal gels \cite{cipelletti3,jain,gordon,jacob,bonn,abou}, aqueous clay suspensions \cite{knaebel,ruzicka,angelini,shahin10,shahin12}, colloid-polymer mixtures \cite{pham04,pham08} and nano-emulsions \cite{gao15}, thus revealing its universal character.

The experimental measurement of the kinetics of aging requires monitoring the dependence of specific physical properties on the evolution (or ``waiting'') time $t$, by means of pertinent experimental techniques. For example, the aging of a glass- or gel-forming material might be represented by the $t$-dependence of the non-equilibrium structure factor $S(k;t)$, and of the collective and self intermediate scattering functions (ISFs) $F(k,\tau;t)$ and  $F_S(k,\tau; t)$. These properties can be accessed, for example, by means of confocal microscopy  \cite{08LuWeitz}, differential dynamic microscopy \cite{10CerbinoTrappe}, x-ray photon correlation spectroscopy \cite{ruta},  and light  \cite{BrambillaCipelletti} or neutron \cite{ruzicka,angelini} scattering. 

Important aspects of the same information, however, are also encoded in the aging of the viscoelastic properties of a given system, more directly amenable through rheometric techniques \cite{09ChristopoulouPetekidisPhilTrans,mewiswagnerbook21}. In fact, the application of rheological concepts and their experimental determination impacts most kinds of soft materials at equilibrium \cite{10ChenJanmeyYodh} and out of equilibrium  \cite{21SumanWagner}. The non-equilibrium rheological or viscoelastic properties of a given system may be represented by the so-called total shear relaxation function $G(\tau;t)$, from which other physical observables can be obtained (e.g. dynamic shear viscosity, complex-valued loss and storage modulus, etc). 
%which is also know experimentally as total viscosity. 
%complex-valued dynamic shear modulus $G(\omega;t)$, whose  real and imaginary parts, $G'(\omega;t)$  and $G''(\omega;t)$, are the ``elastic''  and ``loss''  frequency-dependent moduli. $G(\omega;t)$ is related with the dynamic  shear viscosity $\eta(\omega;t)$ by the relationship $G(\omega;t) =  i\omega \eta(\omega;t)$. 
%Either of these  properties determines the macroscopic stress induced in a viscoelastic liquid upon the application of a small-amplitude oscillatory shear strain of frequency $\omega$ \cite{ferry}. The function $\eta(\omega;t)$ is the  Fourier-Laplace transform of the total shear stress relaxation function $\eta(\tau;t)$. 

From a theoretical perspective, to better control the rheological behavior of colloidal or macromolecular solutions  -- either, for industrial or academic purposes -- one needs a deeper fundamental understanding of its microscopic physical origin, ideally based on theories capable of explaining non-equilibrium viscoelastic properties in terms of the microscopic particle-particle interactions. One possibility \cite{17ZhangLiu}, is to first identify the link between these effective inter-particle forces and the structural and dynamical properties ($S(k; t)$, $F(k,\tau; t)$, $F_S(k,\tau; t)$,\ ...), and then identify (or build) a second link, now between such properties and the system's viscoelastic behavior, represented, for example, by the zero shear viscosity (ZSV) $\eta(t)\equiv\int_0^\infty d\tau G(\tau;t)$. 

Regarding the connection between inter-particle forces and non-equilibrium structure and dynamics, let us recall that, for equilibrium systems, this link is well established and occurs in two steps. In the first of them, approximate statistical mechanical theories (integral equations, density-functional theory, etc. \cite{mcquarrie,hansen}) allow us to determine the equilibrium structure factor $S(k)$ for a given pair interaction potential. Each of these equilibrium theories corresponds to a specific approximation at the level of the second functional derivative $\mathcal{E}[\mid \mathbf{r}-\mathbf{r}'\mid ; n,T] \equiv  \left[ \delta^2 (\mathcal{F}[n,T]/k_BT)/\delta n(\mathbf{\mathbf{r}})\delta n(\mathbf{\mathbf{r}'}) \right]$ of the Helmholtz free energy density-functional $\mathcal{F}[n;T]$, since the equilibrium condition for $S(k;t)$ is the Ornstein-Zernike (OZ) equation $nS(k)=1/\mathcal{E}(k;n,T)$ \cite{orlando}, where $n=N/V$ and the thermodynamic stability function $\mathcal{E}(k;n,T)$ is the Fourier transform of $\mathcal{E}[r; n,T]$ \cite{evans}. In the second step, equilibrium dynamical theories such as mode coupling theory (MCT) \cite{goetze1,goetze2}, the nonlinear Langevin equation theory \cite{2014MirigianScheizer}, or the self-consistent generalized Langevin equation (SCGLE) theory \cite{scgle1,scgle2,todos1}, employ $S(k)$ as an input to yield the \emph{equilibrium} collective and self dynamics, described by the ISFs  $F(k,\tau)$ and $F_S(k,\tau)$. 

A non-equilibrium version of this link between particle interactions and time-dependent structure and dynamics is provided by the NESCGLE theory. The  fundamental principles upon which this theory was built, were laid down and explained in detail in Refs. \cite{nescgle1,nescgle3}. However, a revised and updated presentation can be found  in a recent publication \cite{noneqOMtheory}. According to this theory, the first step is essentially the same as in equilibrium, and consists of an approximation at the level of the thermodynamic properties $\mathcal{F}[n;T]$ or $\mathcal{E}(k;n,T)$. This information, however, is  processed in a fundamentally different manner: for a fluid that is instantaneously quenched at time $t=0$  to a final state point $(n,T)$, the non-equilibrium structure factor $S(k;t)$ is now governed by the time-evolution equation 
\begin{equation}
\frac{\partial S(k;t)}{\partial t} = -2k^2 D^0
b(t)n\mathcal{E}(k;n,T) \left[S(k;t)
-1/n\mathcal{E}(k;n,T)\right], 
\label{relsigmadif2pp}
\end{equation}
which, thus, can be considered as a non-equilibrium extension of the OZ equation \cite{orlando}. In Eq. \eqref{relsigmadif2pp}, $D^0$ is the short-time self-diffusion coefficient, and the  time-dependent mobility function  $b(t)\equiv  D_L(t)/D_0$ is the instantaneous, normalized,  ``long-time'' self-diffusion coefficient $D_L(t)$. 

Eq. \eqref{relsigmadif2pp} thus links the inter-particle forces, contained in $\mathcal{E}(k;n,T)$, with the time-evolving structure, represented by $S(k;t)$. In contrast with the equilibrium case, however, within the NESCGLE theory the kinetic parameters such as the mobility function $b(t)$, are also considered state functions \cite{22innerclocks}, whose ``kinetic equations of state'' must also be determined. In the present case, this is embodied in the fact that $b(t)$ is in reality a functional of $S(k;t)$, determined by a  closed set of equations, whose solution yields  simultaneously $b(t)$, $S(k; t)$, $F(k,\tau; t)$, and $F_S(k,\tau; t)$. This set of NESCGLE equations (summarized  by Eqs. (2.31)-(2.35) of Ref. \cite{orlando}), 
provides precisely the link between inter-particle forces %(encrypted in the thermodynamic input $\mathcal{E}(k;n,T)$) 
and the non-equilibrium structural and dynamical properties.  

Let us now discuss the connection between $S(k; t)$, $F(k,\tau; t)$, and $F_S(k,\tau; t)$ and the non-equilibrium viscoelastic properties represented by the relaxation function $G(\tau;t)$. For equilibrium systems, a simple intuitive approach was introduced by Mason and Weitz as the basis of the experimental technique referred to as ``passive rheology''  \cite{masonweitz}, which assumes a generalized Stokes relation $G (\tau) \approx \zeta(\tau)/ 3\pi \sigma$ between $G(\tau)$ and the memory function $\zeta(\tau)$ of the generalized Langevin equation $M(d\mathbf{v}(\tau)/d\tau)=-\int_0^\tau \zeta(\tau-\tau') \mathbf{v}(\tau') d \tau' + \mathbf{f}(\tau)$ of a tracer particle of diameter $\sigma$. By simply extrapolating this equilibrium assumption to the non-equilibrium domain, we have that  $G (\tau;t) \approx \zeta(\tau;t)/ 3\pi \sigma$. If, in addition, we use the fact that $\zeta(\tau;t)$ is also provided as an output of the NESCGLE equations (see Eq. (2.33) of Ref. \cite{orlando}), we immediately obtain an approximate manner to calculate the non-equilibrium \emph{linear} viscoelasticity from first principles, i.e., for any given  inter-particle  interactions.

Here, however, we shall discuss a more fundamental expression for $G (\tau;t)$, which will be written as a functional of $S(k; t)$ and $F(k,\tau; t)$. For this, let us write
\begin{equation}
 G(\tau;t)= 2\delta (\tau) \eta_s + \Delta G (\tau;t)
\label{noneqmctbanchio1}
\end{equation}
where $\eta_s= k_BT/3\pi\sigma D^0$ is the ``short-time'' (or ``infinite-frequency'') viscosity, and with $\Delta G(\tau;t)$ being the ``non-ideal'' contribution due to the inter-particle forces, which can be written as \cite{orlando,orlando2},
\begin{equation}
 \Delta G (\tau;t)= \frac{k_BT}{60\pi^2} \int_0^\infty dk\ k^4 \left[ \frac{\partial \ln{S(k;t)}}{\partial k} \right]^2\left[  \frac{F(k,\tau;t)}{S(k;t)} \right]^2 .
\label{noneqmctbanchio2}
\end{equation}
This equation should immediately be identified with an analogous equilibrium expression, originally derived by Geszti \cite{geszti} for atomic fluids using MCT, and extended by N\"agele and Bergenholtz \cite{nagelebergmixt} to describe the linear viscoelasticity of colloidal mixtures, which was later applied to compare with equilibrium simulations \cite{banchionageleberg}. Moreover, the stationary solutions of Eqs. \eqref{relsigmadif2pp}-\eqref{noneqmctbanchio2} allow us to describe, under a unique theoretical framework, the rheological behavior of two mutually-exclusive sets: the well-understood universal catalog formed by all the states that maximize entropy, and the complementary universal catalog of non-equilibrium amorphous states of matter, such as gels and glasses \cite{orlando2}.

Of course, the validity of these claims must be tested by contrasting the predictions of Eq. \eqref{noneqmctbanchio2} with pertinent experimental or simulated data. To this endeavor, let us refer to the experimental work reported in Ref. \cite{sam}, which concerns the glassy behavior of suspensions of SiO$_2$ nanoparticles tethered with polyethylene glycol (PEG) chains and suspended in a PEG host. Specifically, here we shall focus on the results of Fig. 5(a) of this reference, which describe the dependence of the ZSV of such PEG-SiO$_2$/PEG suspensions on the volume fraction $\phi$ of nanoparticle cores, for three different diameters ($D_1=10$nm, $D_2=16$nm and $D_3=24$nm) and approaching a transition from low-viscosity-liquid behavior to a glassy (``wax-like'') regime. 

For methodological convenience, and to better emphasize the relevance of Eq. \eqref{noneqmctbanchio2} in analyzing these experimental data, we first applied the SCGLE theory, which in most aspects is equivalent to MCT \cite{voigtmann}. As already mentioned, these theories only require the function $\mathcal{E}(k;n,T)$ as an input (or, equivalently, the structure factor $S(k)$) to yield the ISFs $F(k;\tau)$ and $F_S(k;\tau)$ which, in turn, allow to obtain the equilibrium value $\eta^{eq}(n,T)$ of the ZSV \cite{nagelebergmixt}. To simplify the determination of this input, we shall rely here on a model system that provides a reasonable representation of the effective interactions among the suspended PEG-SiO$_2$ particles, namely, a ``hard-sphere plus repulsive Yukawa'' (HSRY) liquid, whose pair potential $u(r)$ is composed by a hard-sphere contribution representing the excluded volume between SiO$_2$-cores, plus a repulsive Yukawa contribution that represents the soft repulsive (``bumper'') effect of the layer of tethered PEG chains. The total pair potential of this model system reads: 
\begin{equation}
u(r)= 
\begin{cases}
\infty & r < \sigma \\
\epsilon \frac{\exp[-z(r/\sigma -1)]}{(r/\sigma )} & r\geq\sigma,
\end{cases}
\label{yukawa}
\end{equation} 
where $\sigma$ is the hard-core diameter (i.e. $\sigma=D$), $\epsilon$ is the magnitude of the repulsive Yukawa tail at contact, and $z^{-1}$ its decay length (in units of $\sigma$). For given $\sigma$, $\epsilon$, and $z$, the state space of this system is spanned  by the number density $n$ and the temperature $T$, written in dimensionless form as, respectively, the volume fraction $\phi= \pi n\sigma^3/6$ and the dimensionless  temperature $T^{\ast}\equiv k_B T /\epsilon$. 

\begin{figure}[ht!]
\includegraphics[scale=0.25]{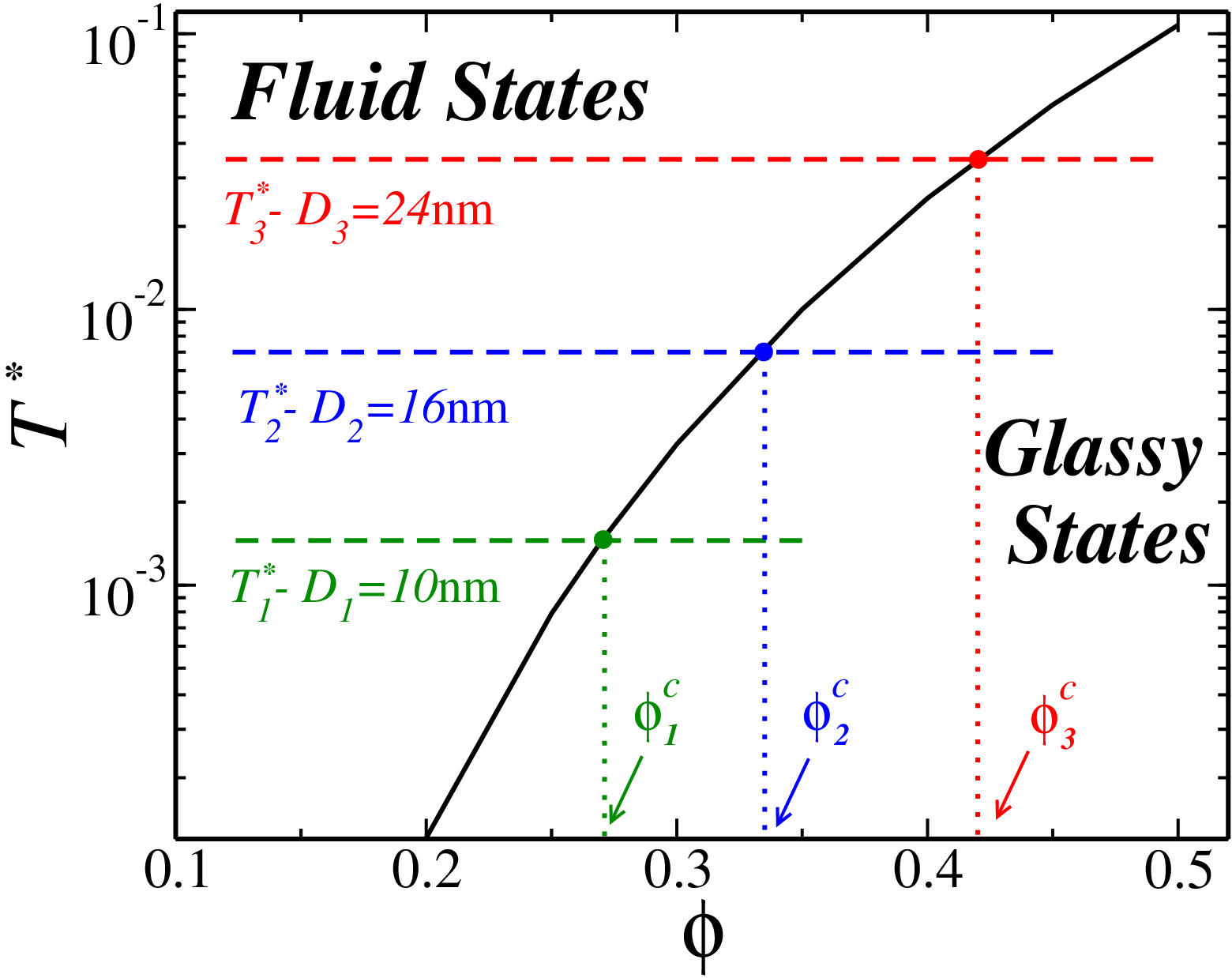}
\caption{Glass transition (GT) diagram predicted by the SCGLE theory for the HSRY system with $z=20$. The solid line partitions the $(\phi,T^*)$-space in two qualitatively different domains: one of ergodic fluid states ($\eta^{eq}(\phi,T^*)$-finite), and one of non-ergodic glassy states ($\eta^{eq}(\phi,T^*)$-infinite). The dashed horizontal lines highlight the three isotherms along which we determined the ZSV using Eq \eqref{noneqmctbanchio2} to compare theory and experiments (see the text).}
\label{fig1}
\end{figure}

Hence, by solving the Ornstein-Zernike equation combined with the hypernetted-chain closure for the HSYR system, we can determine the input $\mathcal{E}(k;\phi;T^*)$, which, combined with the SCGLE and Eq. \eqref{noneqmctbanchio2}, yields the function $\eta^{eq}(\phi,T^*)$ for any given state point. A systematic exploration across the whole parameters space, in turn, leads to the identification of two qualitatively distinct domains in the ($\phi,T^*$)-plane, corresponding to a region of (ergodic) fluid states, where $\eta^{eq}(\phi,T^*)$ is finite, and a region of (dynamically arrested) glassy states, where $\eta^{eq}(\phi,T^*)$ becomes infinite. The boundary between these two regions defines the so-called ideal glass transition (GT) line of the HSRY model,  highlighted in Fig. \ref{fig1} by the black solid curve, corresponding to the case $z=20$ and which will serve as a reference for the upcoming discussion. 

The technicalities involved in both, the determination of the function $\mathcal{E}(k;\phi;T^*)$, and in carrying out the following quantitative comparisons between theory and experiments, are discussed in the supplementary material. There, for example, we explain the rationale to employ three different isotherms $T_1^*,T_2^*$ and $T^*_3$ (dashed horizontal lines in Fig. \ref{fig1}) to calculate the relative viscosity $\eta_r^{eq}(\phi)\equiv \eta^{eq}(\phi;T^*=T_i^*)/\eta_s$ ($i=1,2,3$) of the HSRY system, in order to represent the packing-fraction sweeps at different $D_i$ considered in Ref. \cite{sam}. Notice that these isotherms intersect the GT line at the critical volume fractions $\phi^c_1=0.27,\phi^c_2=0.335$ and $\phi^c_3=0.42$.  

\begin{figure}[h]
\includegraphics[scale=0.25]{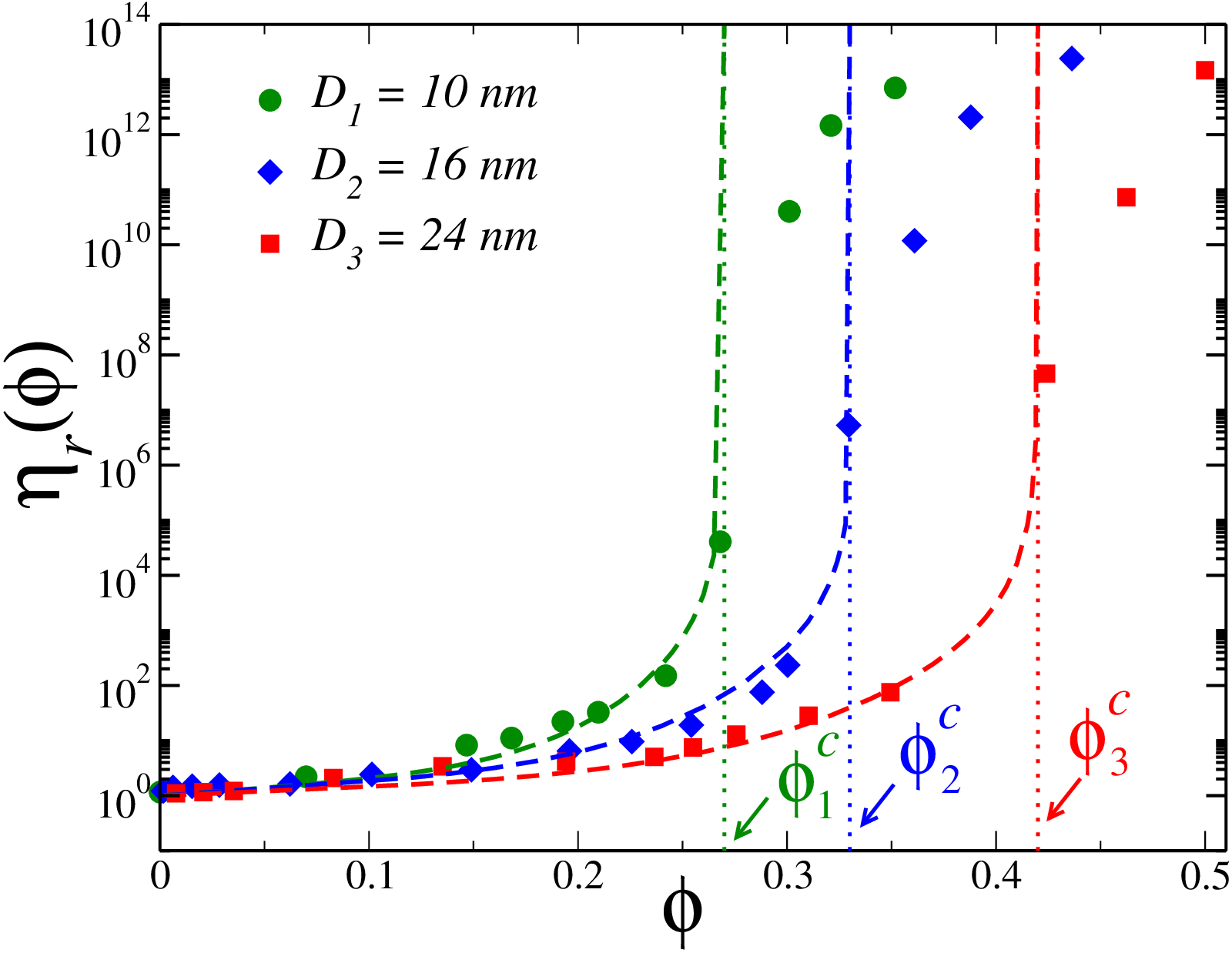}
\caption{Experimental data of Ref. \cite{sam} (solid symbols) for the relative zero-shear viscosity $\eta_r\equiv\eta/\eta_s$, plotted as a function of the volume fraction $\phi$ of SiO$_2$ cores, for three different diameters (as indicated). The dashed lines correspond to theoretical results obtained with the SCGLE for $\eta^{eq}_r(\phi)\equiv\eta^{eq}(\phi,T^*=T^*_i)/\eta$ along the three isotherms shown in Fig. \ref{fig1} (see the text).}
\label{fig2}
\end{figure}

In Fig. \ref{fig2} we reproduce the experimental data of Ref. \cite{sam} for $\eta_r(\phi)$, plotted as a function of $\phi$ for three SiO$_2$-core diameters, as indicated (solid symbols). Focusing on the results for $D_1=10$nm (\textcolor{ao(english)}{$\CIRCLE$}), for example, one notices that the viscosity of the PEG-SiO$_2$/PEG suspension shows relatively small variations within the interval $0<\phi\leq0.2$. For larger concentrations, instead, a progressively stronger growth of the viscosity is observed with increasing packing fraction, where $\eta_r(\phi)$ rises approximately 4 orders of magnitude as $\phi$ approaches the characteristic value $\phi^{exp}_{D_1}\approx0.27=\phi_1^c$. Remarkably, this scenario resembles qualitatively that observed in colloidal hard-sphere suspensions near the GT point \cite{poon,cheng}, at which both MCT and SCGLE predict a divergence of the ZSV \cite{nagelebergmixt}.  As for HSs \cite{BrambillaCipelletti}, in the present case no divergence is observed for $\phi>\phi^{exp}_{D_1}$, but the onset of a transition to a ``wax-like" solid regime, marked by a sharp increase in $\eta_r(\phi)$ by a factor of $10^7$ or more, where this quantity also shows a different growth-rate with increasing packing fraction. The relatively small values of $\phi$ at which this transition is found suggest that the tethered PEG chains induce larger excluded volume effects among the SIO$_2$ cores, thus influencing strongly the viscous behavior of the suspension even at moderate particle concentrations \cite{sam}.

Overall, the same features are observed in $\eta_r(\phi)$ for larger core diameters, $D_2=16$nm (\textcolor{blue}{$\Diamondblack$}) and $D_3=24$nm (\textcolor{red}{$\blacksquare$}), but one notices that the characteristic volume fractions $\phi^{exp}_{D_2}\approx0.335=\phi^c_2$ and $\phi^{exp}_{D_3}\approx0.42=\phi^c_3$ that signal the steep jump in the relative viscosity increase progressively with the particle diameter, thus indicating that the additional contributions to the excluded volume effects produced by the tethered PEG chains contribute the most for the smaller diameter.  

For further reference, in Fig. \ref{fig2} we also display results obtained with the SCGLE theory and Eq. \eqref{noneqmctbanchio2} for the relative viscosity $\eta_r^{eq}(\phi)$ of the HSRY system (dashed lines), calculated along the three isotherms considered in Fig. \ref{fig1} to represent the different core diameters explored in the experiments. The first thing to highlight in these results is their almost perfect agreement with the experimental data for concentrations at-and-below the corresponding critical values $\phi^c_i$, which, according to the SCGLE, mark the onset of a GT in the HSRY liquid. Thus, within the ergodic regime the SCGLE and Eq. \eqref{noneqmctbanchio2} provide a consistent description of the $\phi$-evolution of the ZSV observed in the experimental samples, in spite of the oversimplified model used to represent the direct interactions among the suspended PEG-SiO$_2$ particles. This includes a remarkable quantitative account of the increasingly stronger development of $\eta_r(\phi)$ in the approach to the characteristic values $\phi^{exp}_{D_i}$, and also, the shift of these singular values to larger concentrations upon increasing the core diameter. Therefore, these results for the HSRY model provide additional support to the interpretation that ``\emph{the tethered PEG and SIO$_2$ particles act in concert as an effective hard core, which leads to a higher effective volume fraction of the suspended phase''} \cite{sam}.

This agreement between theory and experiments in the equilibrium regime, however, is to be contrasted with the discrepancy found for packing fractions above $\phi^c_i$, i.e. in the glassy regime where the SCGLE (just as MCT) yields an infinite value for $\eta_r^{eq}(\phi,T)$, as a reminiscence of the divergence of the $\alpha$-relaxation time predicted by this theory at-and-beyond the GT point. Clearly, this idealized physical scenario involving an infinite viscosity is not observed in the experimental results for the PEG-SIO$_2$/PEG suspensions. It is precisely at this point that the relevance of incorporating a kinetic perspective, such as that provided by the NESCGLE theory and Eq. \eqref{noneqmctbanchio2}, becomes crucial. Let us elaborate.   

Both SCGLE and MCT are approaches intrinsically limited to the description of ``nearly arrested" but fully equilibrated liquids. Hence, the phenomenology of transient time-dependent processes such as aging, which is one of the most fundamental fingerprints of the process of amorphous solidification observed in glass-and-gel forming materials, falls completely out of the scope of these equilibrium theories. Let us recall here that a well-known experimental fact is that viscoelastic properties, such as the loss and storage moduli that can be obtained from the relaxation function $G(t;\tau)$, typically show a slow but long-lasting evolution with the observation time $t$ during the formation of non-equilibrium glassy and gelled states \cite{09ChristopoulouPetekidisPhilTrans,jacob,21SumanWagner}. As discussed recently \cite{orlando,orlando2}, the enriched kinetic description provided by the NESCGLE theory and  Eq. \eqref{noneqmctbanchio2} provides a route for the description of this kind of out-of-equilibrium rheological observables, since it allows to determine the full $t$-evolution of the relaxation function $\Delta G(t;\tau)$, from which other properties can be obtained, including the time-dependent ZSV $\eta(t)$. 

In this regard, a meaningful general prediction of the NESCGLE theory and Eq. \eqref{noneqmctbanchio2} is that, for any measurement of the relative ZSV $\eta_r(t;\phi)$ in a glass-forming system, obviously carried out always at a finite waiting time $t=t_0$, the plot of $\eta_r(t=t_0;\phi)$ vs $\phi$ shows two distinct regimes \cite{orlando2}. The first, corresponding to samples that have fully equilibrated within the time $t_0$, and hence, follow the equilibrium plot of $\eta^{eq}(\phi)=\eta(t\to\infty;\phi)$. The second regime, instead, corresponding to samples that, at time $t_0$, have not yet reached equilibrium and, hence, continue aging. From a physical point of view, these predictions indicate that the passage from the low-viscosity equilibrium-liquid regime of the SiO$_2$/PEG suspensions, to the high-viscosity glassy (or ``wax-like'') regime, is not an abrupt transition, but a soft crossover between them, completely free from unobservable divergences. Remarkably, this is precisely the scenario revealed by the experiments of Ref. \cite{sam}.

\begin{figure}[t]
\includegraphics[width=.35\textwidth, height=0.32\textwidth]{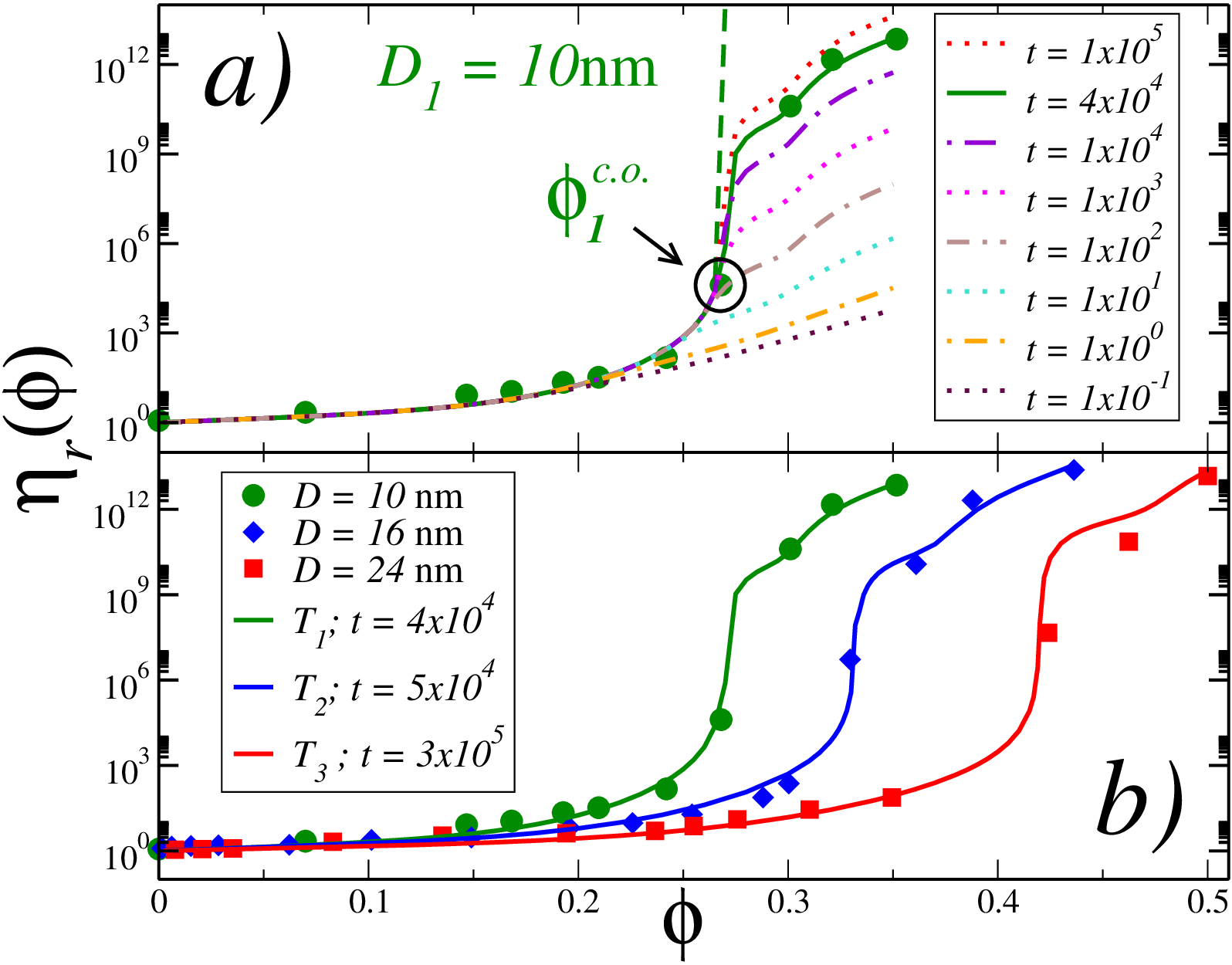}
\caption{a) Sequence of snapshots of the plot $\eta_r(t,\phi)$ vs $\phi$ describing the evolution of the relative viscosity of the HSRY system. The solid circles correspond to the same experimental data of Fig. \ref{fig2} for samples with $D_1=10$nm. The solid curve highlights the predictions of Eq. \ref{noneqmctbanchio2} for $\eta_r(t=4\times10^4;\phi)$ (see the text). b) Comparison of the theoretical results obtained for $\eta_r(t,\phi)$ along the three isotherms in Fig. \ref{fig1} at fixed times (as indicated) and the experimental data of Ref. \cite{sam}.}
\label{fig3}
\end{figure}

Thus, let us reexamine the above experimental data, but this time, employing the enriched lens provided by Eqs. \eqref{relsigmadif2pp}-\eqref{noneqmctbanchio2} for their theoretical interpretation. This allows to discuss, not only the $\phi$-dependence of the relative viscosity, but more crucially, its dependence on the waiting time $t$ at which one starts its measurement. Fig. \ref{fig3}a, for example, considers a sequence of snapshots at different $t$ for the plot of $\eta_r(t;\phi)$ vs $\phi$, obtained from a sequence of instantaneous quenches (at $t=0$) covering all the packing fractions along the isotherm $T^*_1$ (corresponding to the case $D_1=10$nm).

As also explained in detail in Ref. \cite{orlando2}, at any fixed time considered, the function $\eta_r(t=t_0;\phi)$ always displays two $\phi$-regimes: one describing the viscosity of samples that have fully equilibriated within the time $t_0$ (and, hence, lie on the dashed curve in Fig \ref{fig3}, corresponding to $\eta^{eq}_r(\phi)$), and another describing samples that have not yet reached equilibrium  at time $t$ and, thus, remain aging. This, in turn, allows to identify a time-dependent crossover concentration $\phi_1^{c.o.}(t)$, separating the two regimes. One notices that, for a waiting time $t=4\times10^4$ (in units of $D_0/\sigma^2$) the crossover value $\phi_1^{c.o.}(t)$ is slightly smaller, almost matching the critical value $\phi_1^c=0.27\approx\phi_{D_1}$ of the experiments. Remarkably, for this selected waiting time, the plot predicted for $\eta_r(\phi;t)$ vs $\phi$ displays an excellent quantitative agreement with the experimental data for all $\phi$ (solid line in Fig. \ref{fig3}a). As shown in Fig. \ref{fig3}b, by performing the same systematic exercise for the samples with $D_2=16$nm and $D_3=24$nm, along the two isotherms $T^*_2$ and $T^*_3$, respectively, we obtain a compelling theoretical description of the glassy behavior observed experimentally in the PEG-SIO$_2$/PEG suspensions.    

The purpose of this letter and the above comparisons, of course, is only to highlight the striking predictive capability of Eq. \ref{noneqmctbanchio2} and the NESCGLE theory which, combined, provide a powerful fundamental tool for the first-principles description and interpretation of rheological properties in out-of-equilibrium amorphous materials. The theoretical framework briefly outlined here and described in full detail in Ref. \cite{orlando2}, is intended to serve as the foundation for more detailed studies in other specific systems and conditions. This includes applying our proposed approach to characterize the non-equilibrium viscoelastic properties in a variety of systems with different inter-particle interactions and relaxation mechanisms, such as colloid-polymer suspensions, highly asymmetric hard-sphere mixtures, or colloidal systems with competing interactions (short-range attractions and long-range repulsions). Additionally, we are convinced that our approach will become instrumental in exploring in more detail the generality and universality of more abstract concepts regarding glassy behavior, such as the emergence of structural and dynamical anomalies \cite{sam1,cipelletti}, the hard-sphere dynamic universality class in the context of rheological properties, and the fundamental connection between the development of the $\alpha$-relaxation time $\tau_{\alpha}(k)$ and the viscoelastic properties. All these possibilities will be addressed in further work.

%%%%%%%%%%%%%%%%%%%%%

%%%%%%%%%%%%%%%%%%%%%

%%%%%%%%%%%%%%

\begin{thebibliography}{99}
%%%%%%%%%%%%%%%%%%%%%

\bibitem{cipellettiramos05} L. Cipelletti and L. Ramos,
\emph{J. Phys.: Condens. Matter} \textbf{17}(2005) R253-R285.

\bibitem{cipelletti2} L. Cipelletti, L. Ramos, s. Manley, E. Pitard, D. A. Weitz,
E.E. Pashkovski, and M. Johansson, \emph{Faraday Discuss.} \textbf{123}, 237
(2003).

\bibitem{lunkenheimer} P. Lunkenheimer, R. Wehn, U. Schneider, and A. Loidl,
\emph{Phys. Rev. Lett.,} \textbf{95} 055702 (2002)

\bibitem{di} X. Di, K. Z. Win, G. B. McKenna, T. Narita, F. Lequeux, S. R.
Pullela, and Z. Cheng, Phys. Rev. Lett. \textbf{106}, 095701
(2011)

\bibitem{bandyopadhyay}
R. Bandyopadhyay, D. Liang, H. Yardimci, D.A. Sessoms, M.A. Borthwick, 
S.G.J. Mochrie, J.L. Harden, and R.L. Leheny, \emph{Phys. Rev. Lett.} 
\textbf{93}, 228302

\bibitem{bissing}
H. Bissig, S. Romer, L. Cipelletti, V. Trappe, and
P. Schurtenberger, \emph{Phys. Chem. Commun.} \textbf{6}, 21 (2003).

\bibitem{hecksher} T. Hecksher, N.B. Olsenb, K. Nissc, and J.C. Dyre
0\emph{J. Chem. Phys}, \textbf{133} 174514 (2010).

%densegels

%\bibitem{joshi14} Y. M. Joshi, Annu. Rev. Chem. Biomol. Eng. 5, 181-202 (2014).

%metllic alloys

\bibitem{das} A. Das, P.M. Derlet, C. Liu, E.M. Dufresne and Robert Maa\ss,
\emph{Nat. Commun.} \textbf{10}, 5006 (2019). https://doi.org/10.1038/s41467-019-12892-1.

\bibitem{ruta} B. Ruta, G. Baldi, G. Monaco and Y. Chushkin
\emph{J. Chem. Phys.} \textbf{138}, 054508 (2013).


%polymers
\bibitem{hutchinson} J. M.Hutchinson, \emph{Progress in Polymer Science}
Volume 20, Issue 4, 1995, 703-760.
\bibitem{bellon} L. Bellon, S. Ciliberto and C. Laroche,
2000 \emph{Europhys Lett.,} \textbf{51}, 551.

\bibitem{baudez01} J.-C. Baudez and P. Coussot, J. Rheol. 45, 1123-1139 (2001).

%gels
\bibitem{cipelletti3} L. Cipelletti, S. Manley, R. C. Ball, and D. A. Weitz,
\emph{Phys.Rev. Lett} \textbf{84}, 2275 (2000).
\bibitem{jain} A. Jain, F. Schulz,  I. Lokteva, L. Frenzel, G. Grübel and
F. Lehmkühler, \emph{Soft Matter} 2020, \textbf{16}, 2864-2872.
\bibitem{gordon} M.M Gordon, Journal of Rheology \textbf{61}, 23 (2017).

%glass
\bibitem{jacob} A. R. Jacob,  E. Moghimi, and  G. Petekidis,
\emph{Physics of Fluids} \textbf{31}, 087103 (2019).
\bibitem{bonn} D. Bonn, H. Tanaka, G. Wegdam, H. Kellay and J. Meunier
\emph{Europhys Lett.} \textbf{45}(1), pp.52-57 (1999)
\bibitem{abou} B. Abou, D. Bonn, and J. Meunier, \emph{Phys. Rev. E}
\textbf{64}, 021510 (2001).


%laponite
\bibitem{knaebel} A. Knaebel, M. Bellour, J. P. Munch, V. Viasnoff, F. Lequeux,
and J. L. Harden, Europhys. Lett. 52, 73 (2000).

\bibitem{ruzicka} B. Ruzicka, E. Zaccarelli, L. Zulian, R. Angelini,
M. Sztucki, A. Moussaid, T. Naryanan and F. Sciortino,
\emph{Nature Materials} \textbf{10}, 56-60 (2011).

\bibitem{angelini}R. Angelini \emph{et al.} Nat. Commun. 5:4049 doi: 10.1038/ncomms5049 (2014).

\bibitem{shahin10} A. Shahin and Y. M. Joshi, Langmuir 26, 4219-4225 (2010).

\bibitem{shahin12} A. Shahin and Y. M. Joshi, Langmuir 28, 5826-5833 (2012).

\bibitem{pham04} K. N. Pham, S. U. Egelhaaf, P. N. Pusey, and W. C. K. Poon, Phys. Rev. E 69,
011503 (2004).

\bibitem{pham08} K. N. Pham, G. Petekidis, D. Vlassopoulos, S. U. Egelhaaf, W. C. K. Poon, and
P. N. Pusey, J. Rheol. 52, 649-676 (2008).

\bibitem{gao15} Y. Gao, J. Kim, and M. E. Helgeson, Soft Matter 11, 6360-6370 (2015).


\bibitem{08LuWeitz} P. Lu, E. Zaccarelli, F. Ciulla, et al., Nature 453, 499 (2008). 

\bibitem{10CerbinoTrappe}    R. Cerbino and V. Trappe, Phys. Rev. Lett. \textbf{100}, 188102 (2008).

\bibitem{BrambillaCipelletti} G. Brambilla et al., Phys. Rev. Lett. \textbf{102}, 085703 (2009).

\bibitem{09ChristopoulouPetekidisPhilTrans} C. Christopoulou, G. Petekidis, B. Erwin, M. Cloitre, And D. Vlassopoulos, Phil. Trans. R. Soc. A  \textbf{367}, 5051 (2009).

\bibitem{mewiswagnerbook21} Theory and Applications of Colloidal Suspension Rheology, edited by N. J. Wagner and J. Mewis (Cambridge University Press, Cambridge, UK, 2021).

\bibitem{10ChenJanmeyYodh} D.T.N. Chen, Q. Wen, P. Janmey, J. Crocker, ans A. Yodh, Ann. Rev. Condens. Matter Phys. 1, 301 (2010).

\bibitem{21SumanWagner} K. Suman and N. G. Wagner, J. Chem. Phys. \textbf{157}, 024901 (2022).

%\bibitem{ferry} J.D. Ferry, {\it Viscoelastic Properties of Polymers}, Wiley, New York (1980).

\bibitem{17ZhangLiu}  Z. Zhang and Y. Liu, Current Opinion in Chemical Engineering \textbf{16}, 48 (2017).

\bibitem{mcquarrie} D. A. McQuarrie. {\em \ Statistical Mechanics}. Harper \& Row (1973).

\bibitem{hansen} J. P. Hansen \& I. R. McDonald. {\em \ Theory of Simple Liquids}. Academic Press Inc. (1976).

\bibitem{orlando} O. Joaqu\'in-Jaime, R. Peredo-Ortiz, M. Medina-Noyola, and L.F. Elizondo-Aguilera, “From equilibrium to non-equilibrium statistical mechanics of liquids”, https://doi.org/10.48550/arXiv.2401.15220 /manuscript submitted to Phys. Rev. E (2024).

\bibitem{evans} R. Evans, Adv. Phys. {\bf 28}: 143(1979).

\bibitem{goetze1}  W. G\"{o}tze, in {\em Liquids, Freezing and Glass Transition},
edited by J. P. Hansen, D. Levesque, and J. Zinn-Justin (North-Holland, Amsterdam, 1991).

\bibitem{goetze2} W. G\"{o}tze and L. Sj\"ogren,
 Rep. Prog. Phys. {\bf 55}, 241 (1992).

\bibitem{2014MirigianScheizer} Mirigian and K. S. Schweizer, J. Chem. Phys., 2014, \textbf{140}, 194506; ibid 194507. 

\bibitem{scgle1}  L. Yeomans-Reyna and M. Medina-Noyola, Phys. Rev. E {\bf
64}, 066114 (2001).

\bibitem{scgle2}  L. Yeomans-Reyna, H. Acu\~{n}a-Campa,
F. Guevara-Rodr\'{\i}guez, and M. Medina-Noyola, Phys. Rev. E {\bf
67}, 021108 (2003).

\bibitem{todos1} L. Yeomans-Reyna, M. A. Ch\'avez-Rojo, P. E. Ram\'{\i}rez-Gonz\'alez, R. Ju\'arez-Maldonado, M. Ch\'avez-P\'aez, and M. Medina-Noyola, Phys. Rev. E {\bf 76}, 041504 (2007).

\bibitem{nescgle1} P. E. Ram\'irez-Gonz\'alez and M. Medina-Noyola,
\emph{Phys. Rev. E} \textbf{82}, 061503 (2010).

\bibitem{nescgle3}  L.E. S\'anchez-D\'iaz, P.E. Ram\'irez-Gonz\'alez,
and M. Medina-Noyola, \emph{Phys. Rev. E} \textbf{87}, 052306 (2013).

%\bibitem{nescgle8} J.M. Olais-Govea, B. Zepeda-L\'opez, L- L\'opez-Flores, and M. Medina-Noyola, \emph{Scientific Reports} \textbf{9}, 16445 (2019).

\bibitem{noneqOMtheory}  R. Peredo-Ortiz, L. F. Elizondo-Aguilera, P. Ram\'irez-Gonz\'alez, E. L\'azaro-L\'azaro, P. Mendoza-M\'endez, and M. Medina-Noyola, Molecular Physics (2023), DOI: 10.1080/00268976.2023.2297991.


\bibitem{22innerclocks} R. Peredo-Ortiz, M. Medina-Noyola, Th. Voigtmann, and L.F. Elizondo-Aguilera, J. Chem. Phys. \textbf{156}, 244506 (2022).

\bibitem{masonweitz} T. G. Mason and D. A. Weitz, Phys. Rev. Lett. \textbf{74}, 1250 (1995).

\bibitem{orlando2} Peredo-Ortiz, R., Joaquín-Jaime, O., López-Flores, L., Medina-Noyola, M., \& Elizondo-Aguilera, L. F. (2024). "Non-equilibrium theory of the linear viscoelasticity of glass and gel forming liquids", arXiv:2402.14242 [cond-mat.soft] /manuscript submitted to Journal of Rheology (2024)

\bibitem{geszti} T. Geszti, J. Phys. C 16, 5805 (1983). 

\bibitem{nagelebergmixt} G. N\"agele and J. Bergenholtz, J. Chem. Phys. \textbf{108}, 9893 (1998).

\bibitem{banchionageleberg} A. J. Banchio,  G.  N\"agele, and J. Bergenholtz, J. Chem. Phys. \textbf{111}, 8721 (1999).

\bibitem{sam} Structure and rheology of nanoparticle-polymer suspensions, S. Srivastava, J. H. Shin and L. A. Archer, Soft Matter, 8, 4097 (2012).

\bibitem{voigtmann} L. F. Elizondo-Aguilera and Th. Voigtmann Phys. Rev. E \text{100}, 042601 (2019).

\bibitem{poon} W. Poon, S.P. M eeker, P.N. Pusey and P.N. Segr\`e , \emph{J. Non Newtonian Fluid Mech.} \textbf{67} (1996) 179-189.

\bibitem{cheng} Z. Cheng, J. Zhu, and P.M. Chaikin, \emph{Phys. Rev. E} \textbf{65}, 041405 (2002)

%\bibitem{wen} Y.H. Wen, J.L. Schaefer, and L.A. Archer* \emph{ACS Macro Lett.} 2015, \textbf{4}, 1, 119–123

\bibitem{sam1} S. Srivastava and .A. Archer and S. Narayanan, Structure and Transport Anomalies in Soft Colloids \emph{Phys. Rev. Lett.} \textbf{110}  148302 (2013).

\bibitem{cipelletti} A.M. Philippe, D. Truzzolillo, J. Galvan-Myoshi, P. Dieudonné-George, V. Trappe, L. Berthier and L. Cipelletti, Phys. Rev. E \textbf{97}, 040601(R) (2018).

%\bibitem{trappe} D. A. Sessoms, I. Bischofberger, L. Cipelletti, and V. Trappe, Phil. Trans. R. Soc. A \textbf{367}, 5013 (2009); arXiv:0907.2329v2.

%\bibitem{nescgle6}  P. Mendoza-M\'endez, E. L\'azaro-L\'azaro, L.E. 
%S\'anchez-D\'iaz, P.E. Ram\'irez-Gonz\'alez, G.P\'erez-\'Angel, and 
%M. Medina-Noyola, \emph{Phys. Rev. E} \textbf{96}, 022608 (2017).

%\bibitem{18RivasLaurati}	R. Rivas-Barbosa, E. L\'azaro-L\'azaro, P. Mendoza-M\'endez, T.Still, V. Piazza, P.E. Ram\'irez-Gonz\'alez, M. Medina-Noyola, and M. Laurati, Soft Matter, \textbf{14}, 5008 (2018). 

%\bibitem{21LiraMckennaPedro}	J. Lira-Escobedo, P. Mendoza-M\'endez, M. Medina-Noyola, G. McKenna, and P.E. Ram\'irez-Gonz\'alez, J. Chem. Phys., \textbf{155}, 014503 (2021)

%\bibitem{22PatyRicardo} P. Mendoza-M\'endez, R. Peredo-Ortiz, E. L\'azaro-L\'azaro, M. Chavez-P\'aez, H. Ruiz-Estrada, F. Pacheco-V\'azquez, M. Medina Noyola, and L. Elizondo-Aguilera, J. Chem. Phys. \textbf{157} 244504 (2022).

%\bibitem{awc} H. C.Andersen, J. D. Weeks and D. Chandler, Phys. Rev. A \textbf{4}, 1597 (1971).

%\bibitem{percusyevick}  J. K. Percus and G. J. Yevick,
%Phys. Rev. {\bf 110}, 1 (1957).

%\bibitem{wertheim} M. S. Wertheim,  J. Chem. Phys.,  \textbf{55}
%4291-4298 (1971).

%\bibitem{verletweiss} L. Verlet and J. J. Weis {\em Phys. Rev. A} {\bf 5},
%939 (1972).

%\bibitem{rizzo} L.F. Elizondo-Aguilera, T. Rizzo and Th. Voigtmann, Phys. Rev. Lett. \textbf{129} 238003 (2022).

%\bibitem{06Likos} C. N. Likos, Soft Matter \textbf{2}, 478 (2006).

%\bibitem{dyneq0}F. de J. Guevara-Rodr\'iguez and M.Medina-Noyola,Phys. Rev. E {68}, 011405 (2003).

%\bibitem{dyneqPRL} P. E. Ram\'irez-Gonz\'alez, L. L\'opez-Flores, H. Acu\~na-Campa, and M. Medina-Noyola,  Phys. Rev. Lett. \textbf{107}, 155701 (2011).

%\bibitem{dyneqPRE1} L. L\'opez-Flores, H. Ru\'iz-Estrada, M. Ch\'avez-P\'aez,  and M. Medina-Noyola, Phys. Rev. E. {88}, 042301 (2013).

%\bibitem{dyneqPRE2} L. L\'opez-Flores, J. M. Olais-Govea, M. Ch\'avez-P\'aez,  and M. Medina-Noyola, Phys. Rev. E. \textbf{103}, L050602 (2021).


%%%%%%%%%%%%%%%%%%%%%
\end{thebibliography}
\end{document}

\title{Supplemental Material \\   Aging of the Linear Viscoelasticity of Glass- and Gel-forming Liquids}

\author{O. Joaqu\'in-Jaime$^{1}$, E. L\'azaro-L\'azaro$^{1}$, R. Peredo-Ortiz$^{1}$, S. Srivastava$^{2}$, M. Medina-Noyola$^{1}$ and L.F. Elizondo-Aguilera$^{3}$}

\affiliation{ Instituto de F\'{\i}sica,
Universidad Aut\'{o}noma de San Luis Potos\'{\i}, \'{A}lvaro
Obreg\'{o}n 64, 78000 San Luis Potos\'{\i}, SLP, M\'{e}xico}
\affiliation{ Department of Chemical  and Biomolecular Engineering, University of California, Los Angeles, Los Angeles, California  90095, United  States}
\affiliation{Instituto de F\'isica, Benem\'erita Universidad Aut\'onoma de Puebla, 
Apartado Postal J-48, 72570 Puebla, Mexico}

\maketitle
\pagestyle{empty}

In this supplementary material (SM) we provide additional information regarding the solution of Eqs. (1)-(3) of the main text, which allow for the theoretical determination of the zero-shear viscosity (ZSV) of the hard-sphere plus repulsive Yukawa (HSRY) system. In addition, we also describe briefly the main details involved in carrying out the quantitative comparison between our theoretical results and the experimental data of Refs. \cite{sam,sam1} below for the relative viscosity of the PEG-SiO$_2$/PEG suspensions. 

\section{STRUCTURE FACTOR $S(k)$ AND THERMODYNAMIC STABILITY FUNCTION $\mathcal{E}(k;n,T)$}

Let us start by discussing the determination of the thermodynamic stability function $\mathcal{E}(k;n,T)$ that appears in Eq. (1) of the manuscript, which plays a fundamental role in the NE-SCGLE description of glassy and gelled states. As explained in Ref. \cite{pre1} (see, for instance, Subsection 3A of this reference) the thermodynamic function $\mathcal{E}(k;n,T)$ is related to the Fourier transform $c(k;n,T)$ of the two-particle direct correlation function, $c(|\mathbf{r}-\mathbf{r}'|;n,T)$, by 
\begin{equation}
n\mathcal{E}(k;n,T)=1-nc(k;n,T)=1/S^{eq}(k;n,T)\nonumber
\end{equation}
where $n=N/V$, and with $S^{eq}(k;n,T)$ being the equilibrium structure factor (SF). 

\begin{figure}[ht!]
\includegraphics[width=0.32\linewidth]{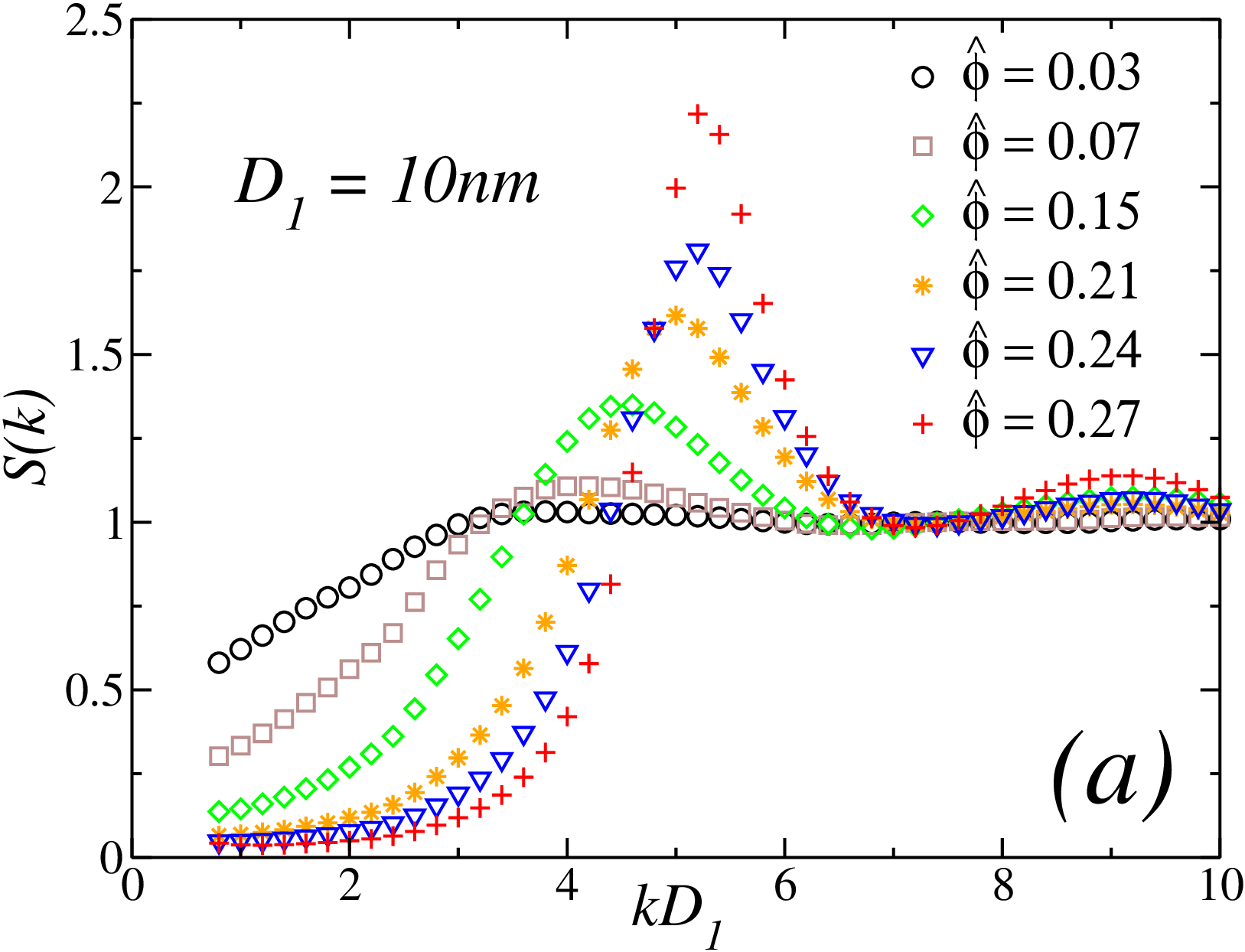}
\includegraphics[width=0.32\linewidth]{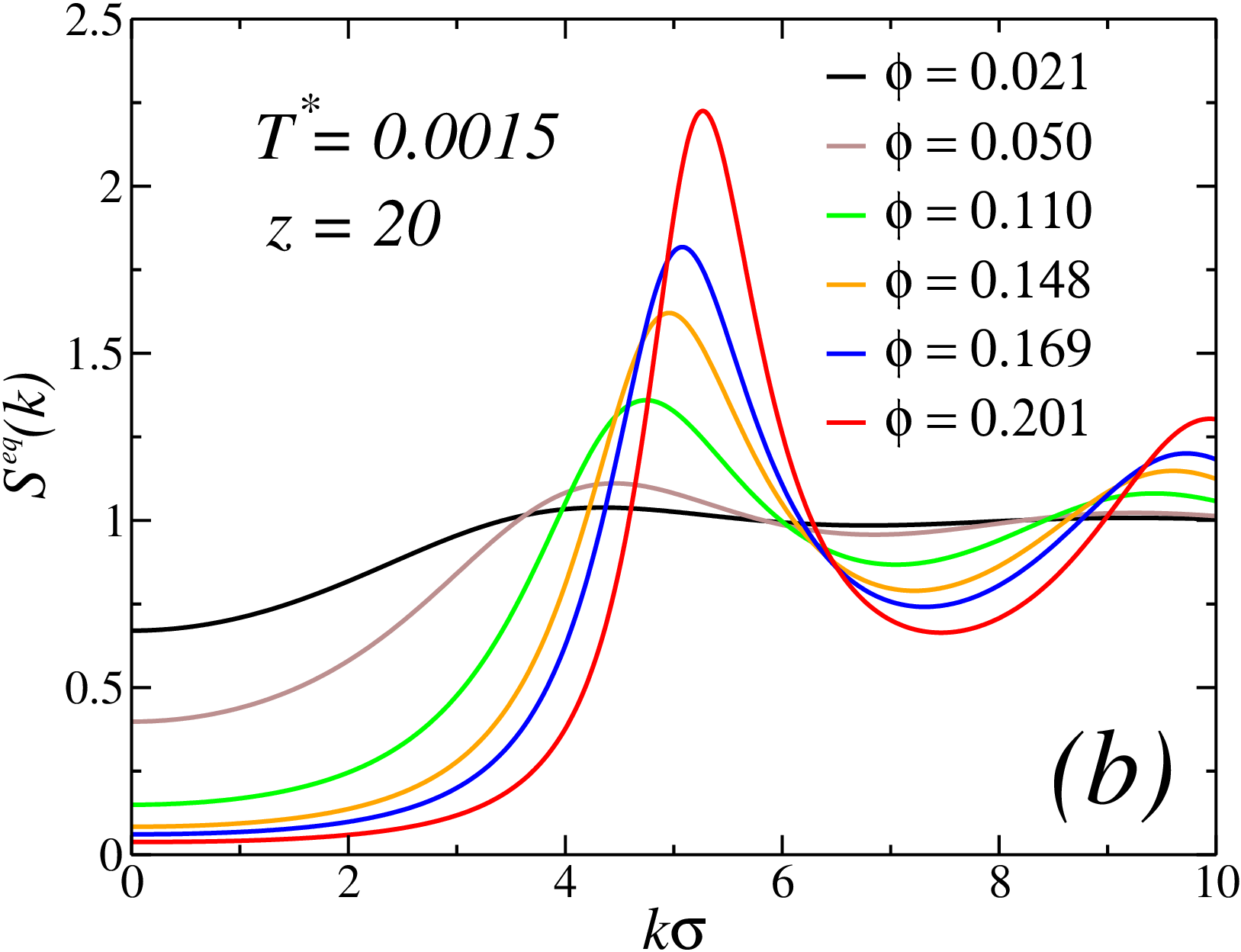}
\includegraphics[width=0.32\linewidth]{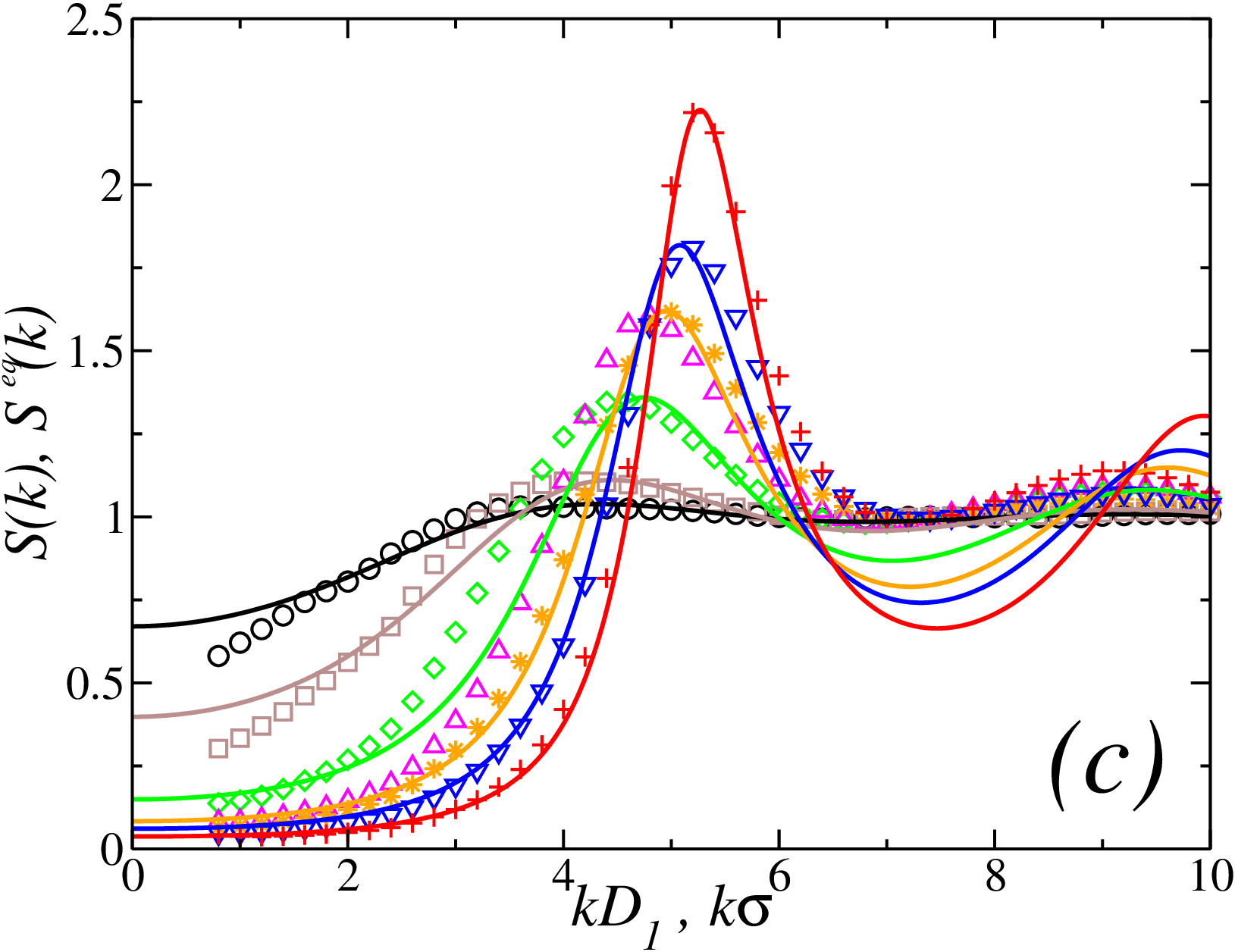}
\caption{(a) Experimental data of Ref. \cite{sam1} describing the $\hat{\phi}$-evolution of the structure factor $S(k)$ for a PEG-SiO$_2$/PEG suspension with SiO$_2$-core diameter $D_1=10$nm. (b) $\phi$-dependence of the equilibrium structure factor $S^{eq}(k)$ of the HSRY fluid, for $z=20$ and along the isotherm $T^*=0.0015$, obtained from the solution of the Ornstein-Zernike equation combined with the hypernetted-chain closure. (c) Quantitative comparison of experimental and theoretical results for $S(k)$ (open symbols) and $S^{eq}(k)$ (solid lines), respectively}
\label{fig1}
\end{figure}

Before describing the procedure to determine $S^{eq}(k)$ for the HSRY system, let us recall the main structural characteristics of the PEG-SiO$_2$/PEG suspensions reported in Refs. \cite{sam,sam1}, where small angle X-ray scattering (SAXS) measurements were carried out to determine the SF, $S(k)$, of three suspensions of  different SiO$_2$-core diameters ($D_1=10$nm, $D_2=16$nm, $D_3=24$nm). Overall, three distinctive features become noticeable in $S(k)$ as the volume fraction $\hat{\phi}$ of SiO$_2$-cores is increased. These are: (i) a gradual development of the primary peak of the SF, (ii) a progressive shift of the primary peak position towards larger $k$-values and, (iii) a gradual decrease in the value of $S(k)$ at small scattering vectors. These features are indicative of both, decreasing interparticle spacing and increasing homogeneity \cite{sam,sam1}. For reference, in Fig \ref{fig1}(a) of this SM we reproduce the experimental data of Fig. 1(a) of Ref. \cite{sam1}, which depicts the $\hat{\phi}$-dependence of the SF of a PEG-SiO$_2$/PEG suspension with SiO$_2$-core diameter $D_1=10$nm.

As discussed in Ref. \cite{sam}, these experimental features (which are also observed for larger diameters $D_2=16$nm and $D_3=24$nm), combined with transmission electron micrographs, provide strong evidence that a PEG-SiO$_2$/PEG suspension can be rigorously thought of as a uniform distribution of nanoparticles in a host polymer. Furthermore, the structural and rheological behavior found experimentally leads to the conclusion that the tethered PEG chains induce additional excluded volume effects among the SiO$_2$ cores, which, collectively, behave as effective hard-core particles, with no evidence of aggregation (see, for instance, Fig. 1 of Ref. \cite{sam}). 
%Furthermore, the structural and rheological behavior found experimentally leads to the conclusion that the tethered PEG chains induce larger excluded volume effects among the SIO$_2$ cores, were the later act collectively, to a good extent, as effective hard-core particles.   

For these reasons, and in order to simplify as much as possible the determination of the function $\mathcal{E}(k;n,T)=[nS^{eq}(k)]^{-1}$, in this work we rely on the HSRY model system, whose equilibrium SF $S^{eq}(k)$ describes qualitatively (and semi-quantitatively) some of the above structural characteristics observed in the PEG-SiO$_2$/PEG suspensions. Specifically, we have solved the Ornstein-Zernike (OZ) equation for the HSRY system defined in Eq. (4) of the main text, using for this the hypernetted-chain closure (HNC) relation, $c(r)=h(r)-\beta u(r)+\ln g(r)$. We recall that the state space of this system is spanned, for fixed $z$,  by the number density $n$ and the temperature $T$, written in dimensionless form as, respectively, the volume fraction $\phi= \pi \rho\sigma^3/6$ and the dimensionless  temperature $T^{\ast}\equiv k_B T /\epsilon$. For further reference, in Fig. \ref{fig1}(b) we display results obtained for $S^{eq}(k)$ for a packing-fraction sweep along the isotherm $T^*_1=0.00145$, fixing the parameter $z$ of the HSRY (the inverse of the decay length) to $z=20$. Here, the numerical values of $T^*$ and $z$ where chosen empirically to obtain the best fit to the experimental data for $D_1=10$nm.

One notices that, except for relatively small quantitative differences in the values for the experimental and theoretical volume fractions ($\hat{\phi}$ and $\phi$, respectively), the solution of the OZ+HNC equation provides a reasonable account of the structural correlations within the interval $k\in[1,6.5]$ (Fig. \ref{fig1}(c)). Beyond this domain, the OZ+HNC and experimental results differ more noticeable, which we attribute to the contributions to the structural correlations produced by the complex interactions among the tethered PEG chains at small distances, which are obviously neglected in the oversimplified model considered here.

Let us mention that, to represent the SF for the larger core diameters ($D_2=16$nm and $D_3=24$nm) we fixed the inverse decay length $z=20$ of the HSRY model, and selected different isotherms ($T^*_2=0.007$ and $T^*_3=0.035$) at larger values of the dimensionless temperature $T^*$. This procedure allowed us to obtain the best fit of the experimental data for $S(k)$, and more importantly, is consistent with the experimental observation that the additional contributions to the excluded volume effects produced by the tethered PEG chains contribute the most for the smaller diameter. Recall that in the HSRY system, these effects are partially represented by the strength parameter $\epsilon$ (see Eq. (4) in the main text).   

\section{GLASS TRANSITION DIAGRAM OF THE HSRY FLUID AND SELECTED ISOTHERMS}

The determination of $S^{eq}(k;\phi,T^*)$ via the OZ+HNC approximation allows for the numerical solution of the SCGLE equations \cite{laura1,laura2} for any given volume fraction $\phi$ and temperature $T^*$, which define the macroscopic state of the HSRY system. Among various dynamical properties, this solution provides the equilibrium value of the intermediate scattering function $F(k,\tau;\phi,T^*)$ which, combined with $S^{eq}(k;\phi, T^*)$, can be employed in Eq. (3) of the manuscript to obtain the equilibrium value $\eta^{eq}(\phi,T^*)$ of the ZSV at any point of the parameters space spanned by $\phi$ and $T^*$.

A systematic exploration over the whole ($\phi,T^*$)-plane, in turn, reveals two main possibilities for the numerical value obtained for $\eta^{eq}(\phi,T^*)$, thus defining two qualitatively distinct domains in this plane, corresponding to: (i) a region of (ergodic) fluid states, where $\eta^{eq}(\phi,T^*)$ yields a finite value; and (ii) a region of (non-ergodic) ideal glassy states, where $\eta^{eq}(\phi,T^*)$ becomes infinite. The boundary between these two regions defines the so-called ideal glass transition GT line, such as that shown in Fig. 1 of the manuscript for the case $z=20$. Of course, the locus of this GT line depends on the specific election of the parameter $z$ of the HSRY system, but one notices that, for any choice, this line always describes a critical volume fraction $\phi^c(T^*)$ that increases monotonically with the scaled temperature. This is illustrated below in Fig. \ref{fig2} for other arbitrarily chosen values of $z$. 

In order to compare quantitatively theoretical results for the ZSV of the HSRY system with experimental data for the PEG-SiO$_2$/PEG suspensions, one must fix a value for $z$, thus defining a GT line. For any choice, the packing fraction sweeps considered in experiments (for diameters $D_1$,$D_2$,$D_3$) might be represented by three isothermal paths where $\phi$ increases. Each specific isotherm $T^*=T^*_i$ was chosen such that it intersect the GT line at a critical value $\phi_i^c\approx\phi^{exp}_{D_i}$, with $\phi^{exp}_{D_i}$ being the characteristic value at which a steep jump in the viscosity of a PEG-SiO$_2$/PEG suspension, with core diameter $D_i$, is observed. From the three diagrams shown in Fig. \ref{fig2}, it becomes clear that any choice of $z$ would provide a distinct isotherm. As already mentioned, however, we stick with the value $z=20$, since it provides a good compromise between $S^{eq}(k)$ and $S(k)$ along three isotherms chosen as just indicated.

\begin{figure}[ht!]
\includegraphics[width=0.5\linewidth]{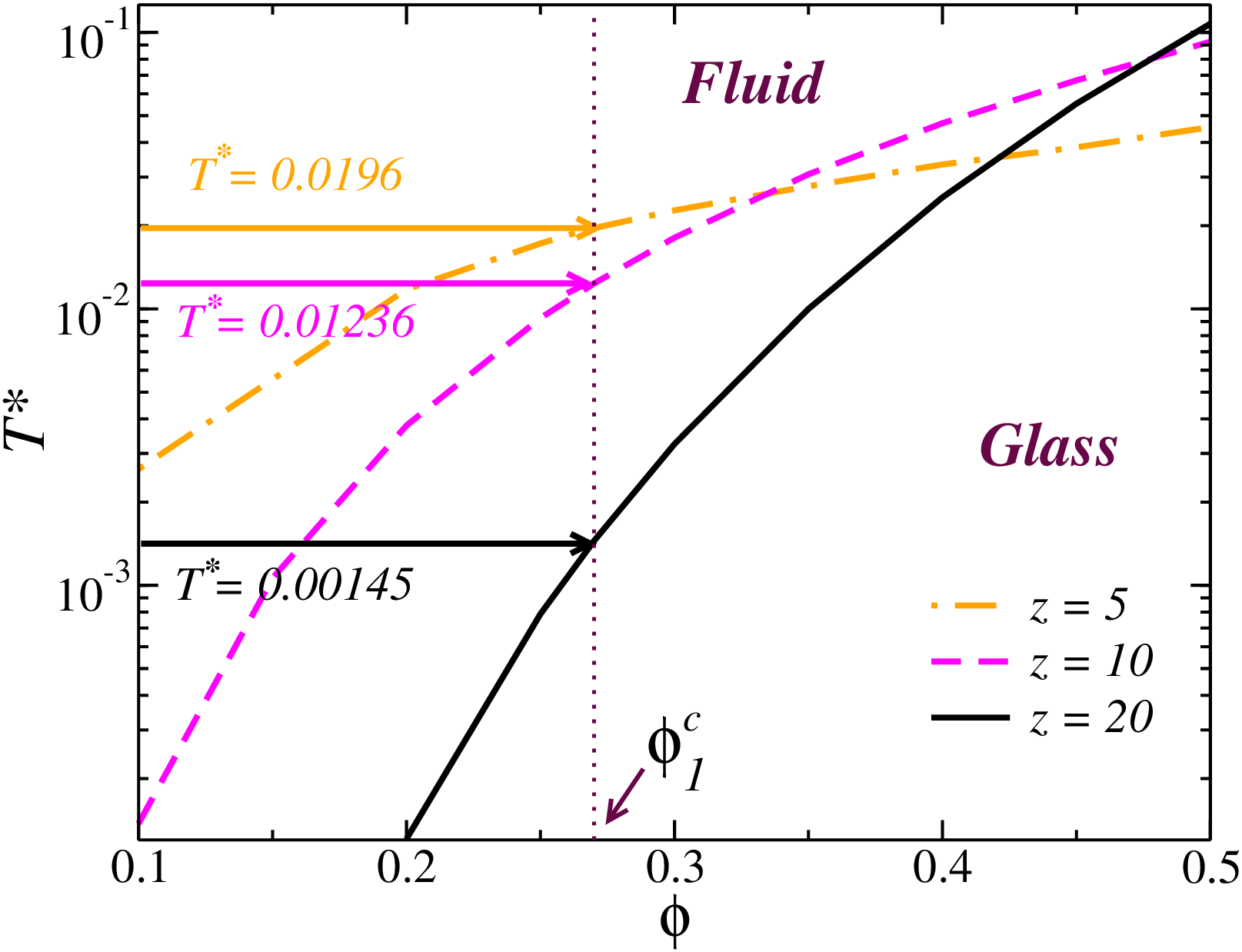}
\caption{Glass transition lines obtained with the SCGLE and Eq (3) in the main text for the HSRY system for $z=5$ (dashed-dotted line), $z=10$ (dashed line) and $z=20$ (solid line). The three horizontal arrows are used to highlight three different isotherms that intersect each of the transition lines at the same critical volume fraction $\phi_1^c=0.27$.}
\label{fig2}
\end{figure}

\section{Brief summary of the NE-SCGLE theory}

For completeness, we briefly describe the essence of the NE-SCGLE theory which, combined with Eq. (3) of the main text, allows for a theoretical description of the  ZSV of the HSRY liquid. The fundamental origin of the NE-SCGLE was laid down in detail in Refs. \cite{pre1, nescgle3}. A revised and updated presentation, however can be found in a recent contribution \cite{todos}. The theory can be summarized by a set of coupled time-evolution equations, whose solution (for $t>0$) describes the irreversible relaxation of an instantaneously quenched liquid. The most central of these equations is Eq. (1) of the manuscript, which describes the waiting-time evolution of the  structure factor $S(k,t)$ for an homogeneous system, instantaneously quenched at $t=0$ from an initial equilibrium state $(\phi_i,T_i^*)$ and towards a final state $(\phi,T^*)$. 

As discussed in the manuscript, the time-evolving mobility function $b(t)$ appearing in the right side of Eq. (1) is defined as $b(t)\equiv D_L(t)/D^0$,
with $D_L(t)$ being the long-time self-diffusion coefficient of the colloidal
particles at evolution time $t$. As explained in Refs. \cite{pre1,nescgle3,todos},
the equation

\begin{equation}
b(t)= [1+\int_0^{\infty}
d\tau\Delta{\zeta}^*(\tau; t)]^{-1}
\label{bdt}
\end{equation}

\noindent relates $b(t)$ with the $t$-evolving and $\tau$-dependent friction
coefficient $\Delta{\zeta}^*(\tau; t)$, given approximately by

\begin{equation}
\begin{split}
  \Delta \zeta^* (\tau; t)= \frac{D_0}{24 \pi
^{3}n_f}
 \int d {\bf k}\ k^2 \left[\frac{ S(k;
t)-1}{S(k; t)}\right]^2  F(k,\tau; t)F_S(k,\tau; t).
\end{split}
\label{dzdtquench}
\end{equation}
Thus, the presence of $b(t)$ in Eq. (1) in the main text, couples $S(k;t)$
with the non-stationary two-time density correlation functions
$F(k,\tau; t)\equiv \displaystyle{\langle \sum_{n,n'}^N \exp(i\mathbf{k}\cdot[
\mathbf{r}_n(t+\tau)-\mathbf{r}_{n'}(t)]) \rangle}$ and $F_S(k,\tau; t)=
\langle \exp[i\mathbf{k}\cdot \Delta\mathbf{r}_T(\tau,t)]\rangle$,
for which the NE-SCGLE also provides time evolution equations. In terms
of their Laplace transforms (LT) $F(k,z; t)$ and $F_S(k,z; t)$, such
equations read,

\begin{gather}\label{fluctquench}
 F(k,z; t) = \frac{S(k; t)}{z+\displaystyle{\frac{k^2D^0 S^{-1}(k;
t)}{1+\lambda (k)\ \Delta \zeta^*(z; t)}}},
\end{gather}

and

\begin{gather}\label{fluctsquench}
 F_S(k,z; t) = \frac{1}{z+\displaystyle{\frac{k^2D^0 }{1+\lambda (k)\ \Delta
\zeta^*(z; t)}}},
\end{gather}

\noindent with $\lambda (k)$ being a phenomenological interpolating
function \cite{gabriel}, given by

\begin{equation}
\lambda (k)=1/[1+( k/k_{c})
^{2}],\nonumber
\label{lambdadk}
\end{equation}
with $k_c$ being an empirically-chosen cutoff wave-vector (here fixed to $k_c=8.2$). This parameter can be employed to calibrate the theory in each specific application (see for instance Sec. III of Ref. \cite{paty}). 

For a given system specified through a pair potential $u(r)$, the solution of the  NE-SCGLE equations provides the $t$-evolution of the functions $S(k;t)$, $b(t)$, $\Delta\zeta^*(\tau;t)$, $F(k,\tau;t)$ and $F_S(k,\tau;t)$, which describe the irreversible relaxation of an instantaneously and homogeneously quenched liquid. More specific details regarding the numerical solution of the NE-SCGLE equations can be found in Ref. \cite{nescgle3}. Hence, one can employ both, $S(k;t)$ and $F(k,\tau;t)$ in Eq. (3) to determine the $t$-evolving relaxation function $G(\tau;t)$, whose $\tau$-integral yields the ZSV $\eta(t)$. 

In practice, to generate the results displayed in Fig. 3 of the manuscript for the time-dependent ZSV $\eta(t)$, one proceeds as follows. For each isotherm $T^*_i$, representing the packing fraction sweeps of the experiments at fixed $D_i$, we choose an initial state ($\phi_i=0.001,T^*=T^*_i$) as a reference, and considered an ensemble of representative HSRY systems, simultaneously subjected to an instantaneous isothermal increase of $\phi$ at $t=0$, starting from the same initial conditions, but with different final packing fraction $\phi$. The set of values of $\phi$, in turn, covering both the ergodic and glassy regimes, i.e. above, at and below the critical value $\phi^c_i$ along the isotherm $T^*_i$.

\section{INCLUSION OF HYDRODYNAMIC INTERACTIONS FOR THE QUANTITATIVE COMPARISON OF  THEORETICAL AND EXPERIMENTAL DATA}

A relevant aspect for the quantitative comparison between our theoretical results and the experimental data of Ref. \cite{sam} concerns to the role of hydrodynamic interactions (HI), which are present in the PEG-SiO$_2$/PEG experimental samples, but that are not taken into account within the framework of SCGLE and NE-SCGLE theories. In this section we outline a simple method to incorporate such hydrodynamic contributions, based on the method proposed in Ref. \cite{medina} to relate the experimental long-time self-diffusion coefficient $D^{exp}_L(\phi)$ of a hard sphere suspension of diameter $\sigma$ at volume fraction $\phi$, with the theoretical prediction $D_L^{theo}(\phi)$ for the ideal version of the same HS suspension in which no hydrodynamic interactions are included. Let us define the dimensionless (or ``reduced'') long-time self-diffusion coefficients $D^{*exp}_L(\phi)$ and  $D^{*theo}_L(\phi)$, respectively, as $D^{*exp}_L(\phi) \equiv D^{exp}_L(\phi)/D^{exp}_L(\phi=0)$ and $D^{*theo}_L(\phi) \equiv D^{theo}_L(\phi)/D_S$. In the first expression, $D^{exp}_L(\phi=0)$ is the self-diffusion coefficient of an isolated particle, which is a constant, related with the viscosity $\eta_0$ of the pure solvent by the Stokes-Einsten relation $D^{exp}_L(\phi=0)=k_BT/3 \pi\sigma \eta_0$. In the second expression, $D_S$ is the short-time self-diffusion coefficient, denoted by $D_0$ in Eq. (1) of the manuscript and in Eqs. (\ref{fluctquench}) and (\ref{fluctsquench}) above. The proposal of Ref. \cite{medina} is to write $D^{*exp}_L(\phi) \equiv D^{exp}_L(\phi)/D^{exp}_L(\phi=0)$ as 
\begin{equation}
D^{*exp}_L(\phi) = \left[\frac{D^{exp}_L(\phi)}{D_S}\right]\left[\frac{D_S}{D^{exp}_L(\phi=0)}\right].
\label{dldsd0}
\end{equation}

The main assumption of  Ref. \cite{medina} is that the first factor on the right side of this equation, which describes the effects of direct interactions, can be approximated by the analogous theoretical property,  which does not include HI, so that using equations \eqref{bdt} and \eqref{dzdtquench}, it is approximated by
\begin{equation}
\left[\frac{D^{exp}_L(\phi)}{D_S}\right] \approx \left[\frac{D^{theo}_L(\phi)}{D_S}\right] = \left[1+\int_0^{\infty}
d\tau\Delta{\zeta}^*(\tau; t)\right]^{-1}.
\label{dldseqtheo}
\end{equation}
On the other hand, in the absence of HI, the short-time self-diffusion coefficient $D_S$ is $\phi$-independent, but the effects of HI on this short-time property leads to a $\phi$ dependence for which Mazur and Geigenm\"uller \cite{mazur} provide a simple expression, namely, 
\begin{equation}
\left[\frac{D_S}{D^{exp}_L(\phi=0)}\right]=\frac{1-\phi}{1+1.5\phi}.\label{etawoh}
\end{equation}
Thus, Eq. (\ref{dldsd0}) can be rewritten as 
\begin{equation}
D^{*exp}_L(\phi) = \frac{1-\phi}{1+1.5\phi}\left[1+\int_0^{\infty}
d\tau\Delta{\zeta}^*(\tau; t)]\right]^{-1}.
%\label{dldsd0}
\end{equation}

These arguments can be translated from the long-time self-diffusion coefficient to the ZSV. For this, let us write 
\begin{equation}
\eta^{HI}_r(\phi) \equiv \left[\frac{\eta (\phi)}{\eta (\phi=0)}\right] = \left[\frac{\eta^\infty(\phi)}{\eta (\phi=0)}\right]\left[\frac{\eta (\phi)}{\eta^\infty(\phi)}\right], 
%\label{dldsd0}
\end{equation}
and by means of the Stokes-Einstein relations: $D^{exp}_L(\phi)=k_BT/3 \pi\sigma \eta(\phi)$ and $D^{exp}_S(\phi)=k_BT/3 \pi\sigma \eta^\infty(\phi)$, let us use the same arguments and approximations employed above for the long-time self-diffusion coefficient, to transform the previous equation in the following result:
\begin{equation}
\eta_r^{HI}\equiv \frac{\eta}{\eta_0}=\frac{1+1.5\phi}{1-\phi}\left(1 + \int_0^\infty \Delta G(\tau)d\tau \right).\label{etawoh}
\end{equation}
Notice that, in the dilute limit, one recovers the well-established results $\eta_r^{HI}\approx1+2.5\phi$. To apply this approximation to the experimental data, to write them in terms of the theoretical calculation of $\Delta G(\tau)$, we must still adapt this method to include the fact that our particles are not strictly hard spheres. The effect of the thickness of the PEG layer on the direct interactions has already been included in the repulsive Yukawa interaction. However, this layer also modifies the hydrodynamic interactions, since the effective  hydrodynamic radius $\sigma_{hd}$ is not identical to the core radius $\sigma$, but is increased by a factor $\lambda>1$,  $\sigma_{hd} \approx \lambda \sigma$, so that the volume fraction in the hydrodynamic factor of the right side of Eq. (\ref{etawoh}) is actually the rescaled volume fraction $\lambda^3 \phi$. 

\begin{figure}[ht!]
\includegraphics[width=0.49\linewidth]{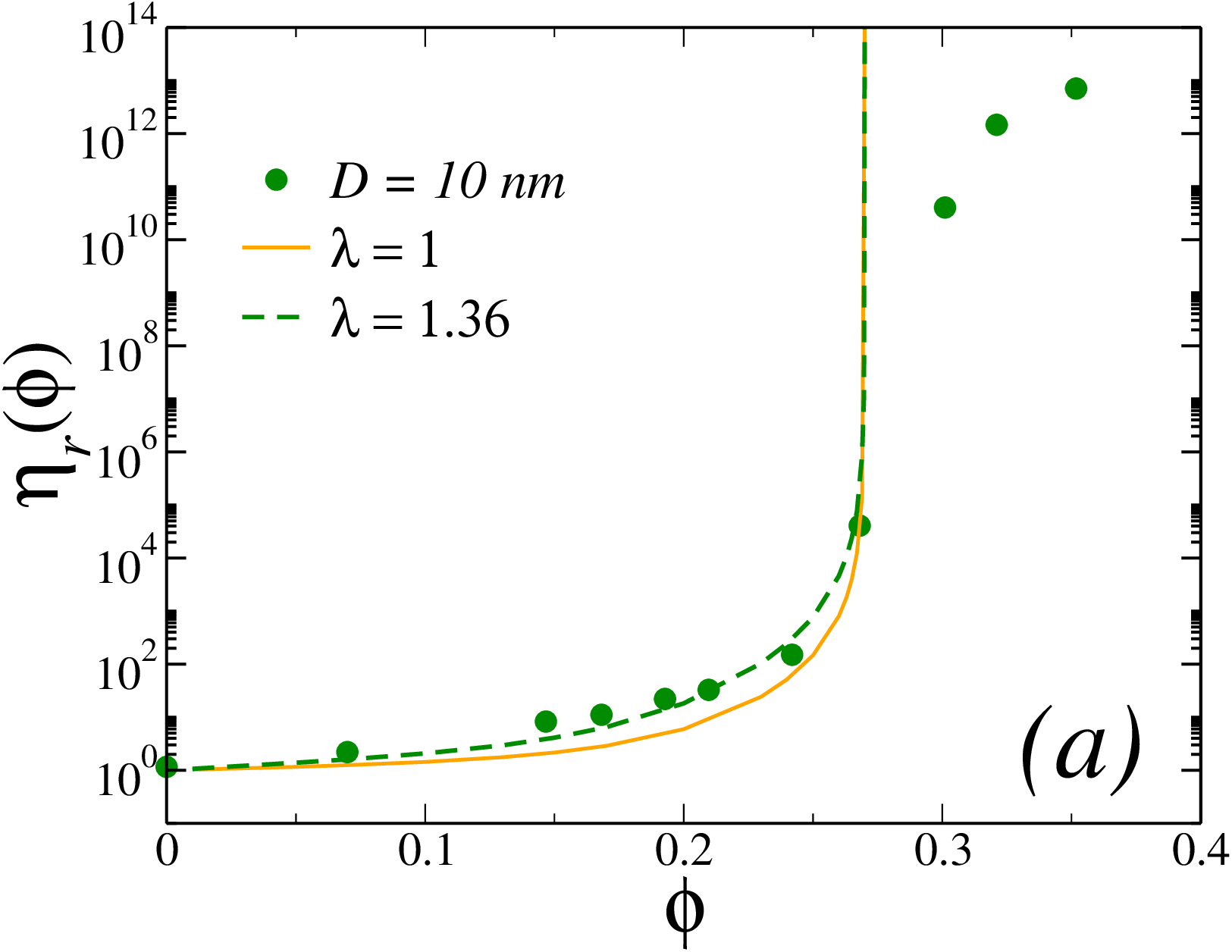}
\includegraphics[width=0.49\linewidth]{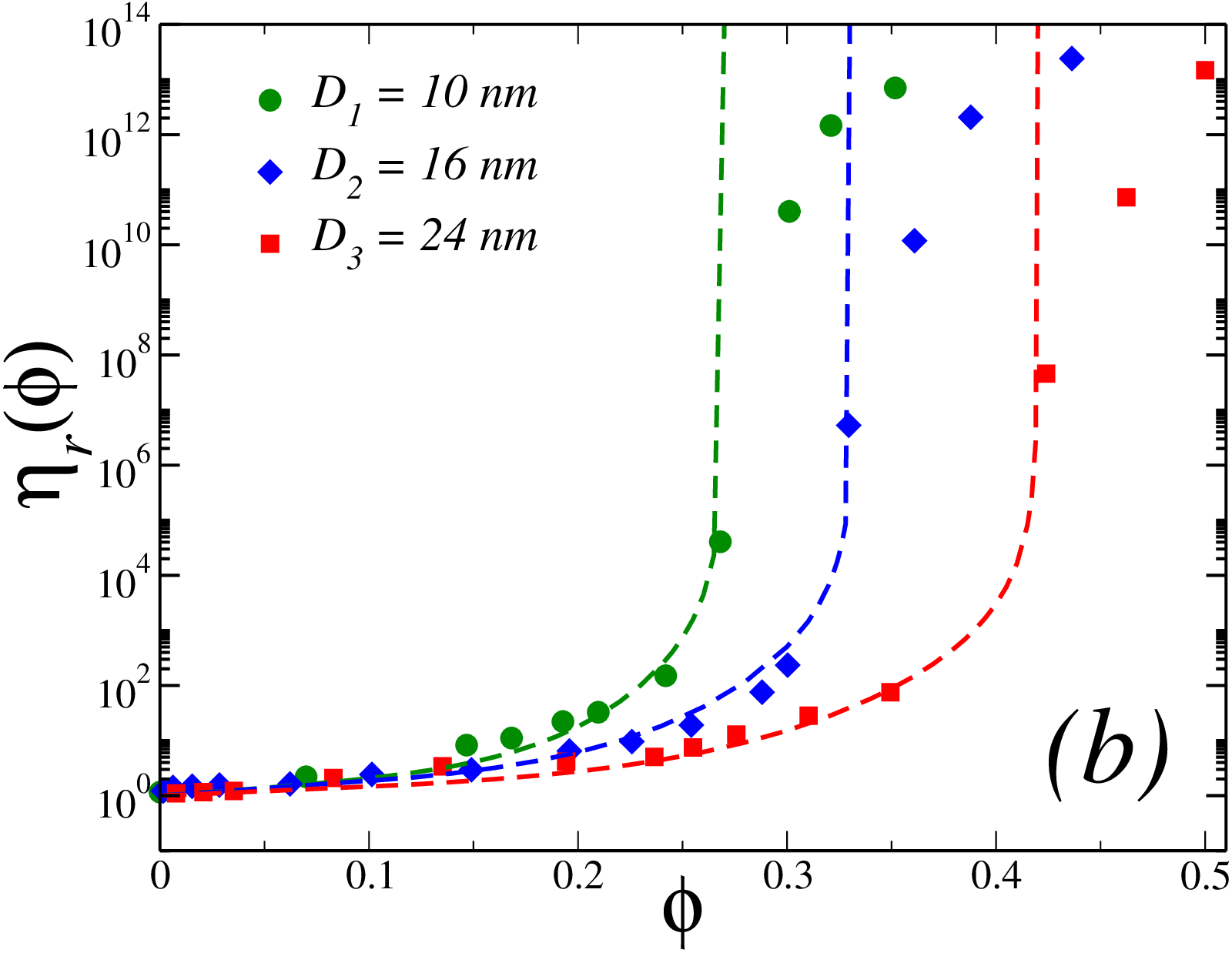}
\caption{(a) Comparison of results obtained with the SCGLE for the ZSV with HI for $\lambda=1.0$ (solid line), $\lambda=1.36$ (dashed line), and the experimental data of Ref. \cite{sam} (solid symbols). (b) Comparison of results obtained with the renormalized SCGLE and experimental data (see the text).}
\label{fig3}
\end{figure}

To assess the validity of the above arguments, in Fig. \ref{fig3}(a) we compare the SCGLE predictions for the ZSVs $\eta^{HI}_r(\phi,T^*_1;\lambda = 1.0)$ (solid line) and $\eta^{HI}_r(\phi,T^*_1;\lambda = 1.36)$ (dashed line) with the experimental data of Ref. \cite{sam} for samples with $D_1 = 10$ nm (solid symbols). Although the results without the factor correction already show essentially the same trends as the experimental results, one notices that the renormalization proposed yields an almost perfect agreement between theory and experiments. Finally, given that the excluded volume effects produced by the tethered PEG become smaller with increasing SiO$_2$ diameter, we might adjust the parameter $\lambda$ for each comparison of theoretical and experimental results. Fig. \ref{fig3}(b) reproduces Fig. 2 of the main text, in which we have tuned the parameter $\lambda$ as $\lambda_1=1.36, \lambda_2=1.29$ and $\lambda_3=1.25$ for, respectively, $D_1=10$nm, $D_2=16$nm and $D_3=24$nm.